\documentclass[letterpaper, 10 pt, conference]{ieeeconf}  

\IEEEoverridecommandlockouts                              
\usepackage[utf8]{inputenc}

\usepackage{amsmath,amsfonts,amssymb,amsthm}
\usepackage{thmtools}
\usepackage{graphicx}
\usepackage[hidelinks]{hyperref}
\usepackage[capitalize]{cleveref}
\usepackage{subcaption}
\usepackage{float}
\usepackage{cite}
\usepackage{balance}
\usepackage{url}

\usepackage{algorithm}
\usepackage{algorithmic}
\usepackage{multirow}
\usepackage{accents}
\newcommand{\ubar}[1]{\underaccent{\bar}{#1}}

\newtheorem{theorem}{Theorem}[section]
\newtheorem{corollary}[theorem]{Corollary}

\crefname{section}{Section}{Sections}
\crefname{theorem}{Theorem}{Theorems}
\crefname{lemma}{Lemma}{Lemmas}
\crefname{table}{Table}{Tables}
\crefformat{equation}{(#2#1#3)}
\crefname{algocf}{Algorithm}{Algorithms}
\Crefname{algocf}{Algorithm}{Algorithms}
\crefname{ALC@unique}{Line}{Lines}

\usepackage{xcolor}
\newcommand{\bluehref}[3][blue]{\href{#2}{\color{#1}{#3}}}%

\usepackage{titlesec}

\title{\LARGE \bf
Learning Verifiable Control Policies Using Relaxed Verification
}

\author{Puja Chaudhury, Alexander Estornell, Michael Everett
\thanks{Authors are with Northeastern University, Boston, MA, USA. e-mail: {\tt\small \{chaudhury.p,estornell.a,\allowbreak m.everett\}\allowbreak@\allowbreak northeastern.edu}}%
}

\begin{document}

\maketitle

\begin{abstract}
To provide safety guarantees for learning-based control systems, recent work has developed formal verification methods to apply after training ends. However, if the trained policy does not meet the specifications, or there is conservatism in the verification algorithm, establishing these guarantees may not be possible. Instead, this work proposes to perform verification throughout training to ultimately aim for policies whose properties can be evaluated throughout runtime with lightweight, relaxed verification algorithms. The approach is to use differentiable reachability analysis and incorporate new components into the loss function. Numerical experiments on a quadrotor model and unicycle model highlight the ability of this approach to lead to learned control policies that satisfy desired reach-avoid and invariance specifications.
\end{abstract}
\textbf{Code:} \bluehref{https://github.com/catplotlib/Learning-Verifiable-Control-Policies-Using-Relaxed-Verification}{https://github.com/catplotlib/Learning-Verifiable-Control-Policies-Using-Relaxed-Verification}.

\section{Introduction}
Accounting for safety requirements remains a challenge in learning-based control, which is essential in applications such as aerospace and autonomous driving. Motivated by this issue, recent work has developed formal verification techniques for neural networks (NNs)~\cite{zhang2018efficient, weng2018towards, xu2020automatic, raghunathan2018semidefinite, tjeng2017evaluating, katz2019marabou, katz2017reluplex, jia2021verifying,vincent2021reachable} and NN-controlled dynamical systems~\cite{sidrane2022overt,chen2023one,everett2021reachability,julian2019reachability,hu2020reach,wang2023polar,ivanov2019verisig,vincent2021reachable,dutta2019reachability, huang2019reachnn, fan2020reachnn, xiang2020reachable,bak2022closed}. For example, these techniques can be used to prove stability, calculate regions of attraction, or estimate forward and backward reachable sets, as part of a comprehensive safety analysis.

However, due to the computational cost of formal verification (e.g., exact verification of input-output properties of ReLU NNs is NP-Complete~\cite{katz2017reluplex}), these techniques are usually applied \textit{after} learning is complete. If the resulting system does not satisfy the safety specifications, naturally, no sound verification algorithm would be able to prove that it does. And even if the resulting system does satisfy safety specs, it may not be tractable to perform complete verification~\cite{vincent2021reachable,wang2021beta,zhang2022safety}, especially for run-time monitoring throughout operation. Meanwhile, the looseness introduced by sound but incomplete (fast) verification algorithms, such as ~\cite{hu2020reach,everett2021reachability,zhang2018efficient,zhang2022safety}, may still prevent obtaining a proof.

\begin{figure}[t]
    \centering
    \includegraphics[clip,trim=0 80 0 20,page=1,width=1\linewidth]{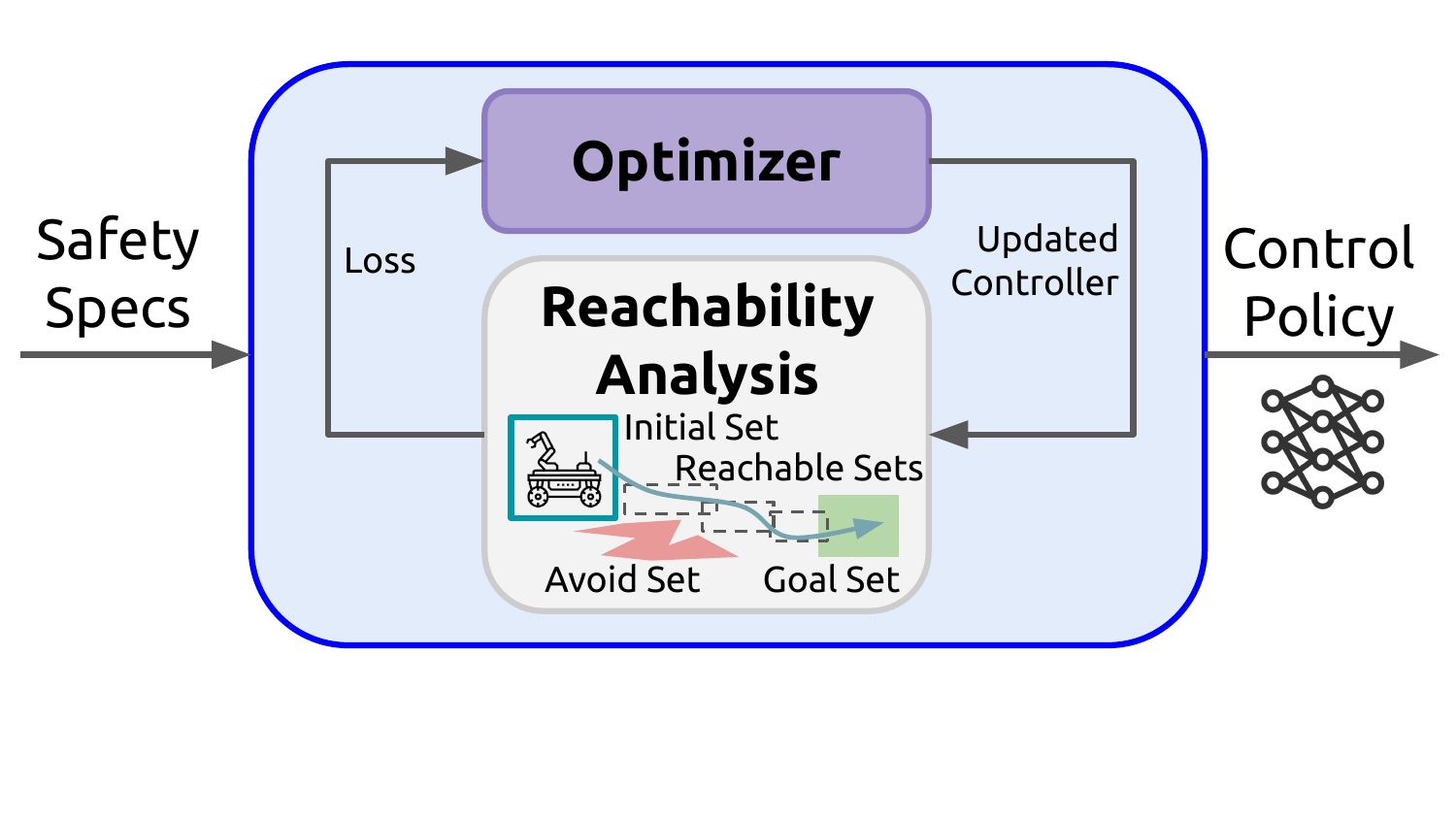}
    \caption{Learning verifiable control policies. The objective is to synthesize a NN control policy guided by a safety specifications. At each training iteration, loss terms based on reachability analysis are used to update the NN parameters.}
    \label{fig:overview}
\end{figure}

Instead, this work focuses on guiding the learning process to meet performance objectives and satisfy safety specifications by performing verification \textit{during} each training iteration.
While this is common in training NNs in isolation (e.g., for image classification~\cite{gowal2019scalable}), relatively fewer have considered synthesizing NN controllers with safety and performance guarantees (e.g., \cite{chang2019neural, sun2020learning, han2020actor, qin2021learning, dai2021lyapunov, dawson2022safe,badings2024learning,yang2024lyapunov,wang2021verification,wang2023joint,wu2024verified}), or more specifically, considered verification while training NN control policies.
For example, several existing methods propose to simultaneously learn a policy and certificate (e.g., \cite{badings2024learning,dawson2022safe,dai2021lyapunov,yang2024lyapunov}) with counterexample-guided training.
A limitation of counterexample-guided training is that it is not obvious which and how many points to sample at each iteration.
Alternatively, \cite{wang2021verification} proposed to add loss terms based on forward reachable sets, with numerical estimates of gradients used to perform policy updates.
\cite{wang2023joint,wu2024verified} further extended this idea using differentiable reachable set estimates for reinforcement learning.
Building on these advances, this work investigates ways to synthesize useful control policies strictly using the differentiable reachable set bounds, without relying on an external reward signal from the environment.

The primary contribution of this work is an approach for guiding NN control policy training to encourage safety specification satisfaction, illustrated in \cref{fig:overview}.
The approach uses CROWN~\cite{zhang2018efficient} to compute reachable set estimates that are differentiable with respect to the controller parameters.
This enables defining reachability-related terms in the loss function, such as robustly reaching a goal region, avoiding obstacles, or forming other invariant sets.
Numerical experiments on unicycle and 6D quadrotor models demonstrate that the policies can be trained quickly and the resulting systems not only satisfy the specifications, but they also can be verified quickly, which is important for run-time monitoring, where the specifications may change throughout operation.

\section{Preliminaries}
\subsection{Problem Statement}

For a state space $\mathcal{X} \subseteq \mathbb{R}^{n_x}$ and control space $\mathcal{U} \subseteq \mathbb{R}^{n_u}$, this paper aims to find a feedback control policy $\pi_{\theta}: \mathcal{X} \to \mathcal{U}$, parameterized by $\theta$, such that a system's closed-loop dynamics satisfy reach-avoid properties.
In particular, denote the discrete-time closed-loop dynamics as:
\begin{equation}
\mathbf{x}_{t+1} = f(\mathbf{x}_t, \pi_\theta(\mathbf{x}_t)), \label{eq:neural_feedback_loop}
\end{equation}
where $\mathbf{x}_t\in\mathcal{X}$ is the system state at time $t$.
When the closed-loop dynamics comprise of NN components (e.g., NN controller, NN dynamics), we will call the system in \cref{eq:neural_feedback_loop} a \textit{neural feedback loop}, which suggests the controller and/or dynamics may be high-dimensional and nonlinear.
It remains challenging to find $\theta$ such that a neural feedback loop satisfies complicated reach-avoid properties (e.g., navigating through an environment with many obstacles).

\subsection{Verification of Neural Networks}

Much of the recent work on neural feedback loops has focused on analysis. That is, given a \textit{trained} control policy with fixed parameters $\theta$, does the policy meet the reach-avoid specifications?
While the \textit{exact} neural feedback loop verification problem can be encoded as a nonlinear program (e.g., mixed-integer program~\cite{sidrane2022overt} when $\pi$ uses ReLU activations), computational limits often motivate sound (but incomplete) verifiers, which leverage convex relaxations to obtain linear or semidefinite programs.

\begin{theorem}[Linear Relaxation-based Perturbation Analysis (LiRPA)]\label{thm:lirpa}
Given a function $g: \mathcal{X}\subseteq\mathbb{R}^n\to\mathcal{Y}\subseteq\mathbb{R}^m$, input set $\mathcal{X}'\subseteq\mathcal{X}$,
and polytope facets $\mathbf{C}\in\mathcal{R}^{c\times m}$, LiRPA calculates $\mathbf{d}\in\mathbb{R}^c$, 
which defines polytope outer bounds on the image $g(\mathcal{X}')\subseteq\{\mathbf{y}\in\mathcal{Y}\,\lvert\,\mathbf{C}\mathbf{y}\leq\mathbf{d}\}$.
\end{theorem}

The details of how LiRPA calculates $\mathbf{d}$ are provided in~\cite{zhang2018efficient}.
For a one sentence summary, LiRPA computes affine bounds on each primitive in the computation graph, $g$, that are guaranteed to hold over the input domain to that primitive, then aggregates all of these affine bounds, then concretizes the bounds from function input to output (in closed-form for $l_p$-ball input domains).
The codebases~\cite{Bunel2023google,Shi2025Verified} support performing LiRPA on computation graphs with a wide range of primitives that appear in neural networks and dynamical systems (e.g., trigonometric functions, affine transformations, ReLU, sigmoid).
Backward CROWN -- which we will refer to simply as CROWN -- is a type of LiRPA.

\subsection{Verification of Neural Feedback Loops}

Moreover, calculating over-approximations of forward reachable sets of a neural feedback loop is a straightforward application of~\cref{thm:lirpa}.
\begin{corollary}[\!\!\cite{zhang2018efficient}, Closed-Loop Reachability Analysis with CROWN]\label{thm:nfl_lirpa}
Given a neural feedback loop $f$, initial state set $\mathcal{X}_t \subseteq \mathcal{X}$, and polytope facets $\mathbf{C}\in\mathbb{R}^{c \times n}$, the system's next state $\mathbf{x}_{t+1}$ must be in the 1-step reachable set, $R_1(\mathcal{X}_t)$, and its  outer bound, $\bar{R}_1(\mathcal{X}_t)$:
\begin{align}
R_1(\mathcal{X}_t) &\triangleq \{ \mathbf{x}_{t+1}\ \lvert\ \mathbf{x}_{t+1} = f(\mathbf{x}_{t}, \pi_\theta(\mathbf{x}_t)\}\\
&\subseteq \{ \mathbf{x}_{t+1}\ \lvert\ \mathbf{C}\mathbf{x}_{t+1} \leq \mathbf{d}\}\triangleq \bar{R}_1(\mathcal{X}_t). \label{eq:reachable_set_bounds}
\end{align}
\end{corollary}

Several recent papers explore the trade-offs between computational cost and bound tightness for estimating reachable sets over $T$ timesteps: either by running \cref{thm:nfl_lirpa} $T$ times iteratively or running \cref{thm:nfl_lirpa} once over a computation graph that contains $T$ copies of $f$ (e.g., \cite{chen2023one,sidrane2022overt,sidrane2024ttt,rober2024constraint}).
This paper will use the former approach.

To check whether the system meets the reach-avoid specifications, common approaches are to (a) calculate the reachable sets explicitly and check for intersection/containment with the avoid/reach sets, respectively, or (b) encode the reach-avoid properties as additional layers at the end of the dynamics and check for feasibility.

In this work, we will let $\mathbf{C}=[\mathbf{I}_{n_x \times n_x},\, -\mathbf{I}_{n_x \times n_x}]^{\top}$ to obtain hyperrectangle bounds on the state vector. Therefore, for $\mathbf{C}\mathbf{x}\leq\mathbf{d}$, let $\mathbf{d}=[\bar{\mathbf{x}}, -\ubar{\mathbf{x}}]^\top$. We will sometimes refer to reachable set bounds $\bar{\mathcal{R}}$ either as a set, as in \cref{eq:reachable_set_bounds}, or by $\bar{\mathbf{x}}$ and $\ubar{\mathbf{x}}$ as the upper and lower (elementwise) bounds on $\mathbf{x}$, respectively.

\section{Approach}

The controller training process incorporates verification results through a loss function with multiple objectives,
\begin{equation}
\mathcal{L}(\theta) = \sum w_{\cdots} \ \mathcal{L}_{\cdots}
\end{equation}
where each term encodes an aspect of system performance:


1) \textit{Goal/Obstacle Overlap}: penalizes the volume of (positional) reachable sets that are not within a region, e.g., around the goal position $\mathbf{x}_g$,
\begin{align}
\mathcal{L}_\text{overlap}&=\sum_{t=0}^{T} \prod_{i=0}^{n_x-1} \text{max}(\text{min}(\bar{\mathbf{x}}_{t,i}, \mathbf{x}_{g,i}+0.5) - \nonumber\\&\qquad\qquad\qquad\quad\text{max}(\ubar{\mathbf{x}}_{t,i}, \mathbf{x}_{g,i}-0.5), 0). \label{eq:loss:goal_overlap}
\end{align}

2) \textit{Goal-Reaching}: drives the system toward the goal region, by penalizing the distance between the center of the goal region, $\mathbf{x}_g$, and the center of each reachable set,
\begin{equation}
\mathcal{L}_\text{goal} =  \sum_{t=0}^{T} \| \frac{\bar{\mathbf{x}}_t+\ubar{\mathbf{x}}_t}{2} - \mathbf{x}_g\|_2.\label{eq:loss:goal}
\end{equation}

3) \textit{Bound Volume}: encourages controller parameters such that the system's reachable sets can be bounded tightly (i.e., small volume) by a \textit{relaxed} verification algorithm,
\begin{equation}
\mathcal{L}_\text{vol} = \sum_{t=0}^{T} \prod_{i=0}^{n_x - 1}\left(\bar{\mathbf{x}}_{t,i} - \ubar{\mathbf{x}}_{t,i}\right).\label{eq:loss:vol}
\end{equation}

4) \textit{Invariance}: encourages establishing an invariant set later in the trajectory, to ensure the system remains sufficiently close to the goal region indefinitely,
\begin{equation}
\mathcal{L}_\text{inv} = \sum_{t=t_\text{inv}}^{T} \| \bar{\mathbf{x}}_{t+1} - \bar{\mathbf{x}}_{t}\|_2 + \| \ubar{\mathbf{x}}_{t+1} - \ubar{\mathbf{x}}_{t}\|_2, \label{eq:loss:inv}
\end{equation}
where $t_\text{inv}$ is the timestep to start encouraging invariance. Without this term, a common issue in using relaxed verification algorithms is that samples from the reachable sets suggest there is likely an invariant set around the goal, but the looseness in the bounds presents difficulty in proving this.

\begin{algorithm}[t]
\caption{Verification-In-The-Loop Training}
\textbf{Input:} Initial states $\mathcal{X}_0$, dynamics $f$, goal $\mathcal{G}$, obstacles $\mathcal{A}_{0:A}$\\
\textbf{Output:} Optimized parameters, $\theta^*$

\begin{algorithmic}[1]\label{alg:verification_in_loop}
\STATE $\theta_0 \leftarrow$ randomly initialized parameters
\STATE optimizer $\leftarrow$ Adam(lr = 0.0001) \cite{kingma2014adam}
\STATE $\bar{\mathcal{R}}_0 \leftarrow \mathcal{X}_0$
\WHILE{not converged}
    \FOR{$t \in \{0, \ldots, T\}$}
        \STATE $\bar{\mathcal{R}}_{t+1} \leftarrow \text{CROWN}(f, \bar{\mathcal{R}}_t)$, where $f$ uses $\pi_{\theta_i}$ \cite{zhang2018efficient} \label{alg:crown}
    \ENDFOR
    \STATE Evaluate $\mathcal{L}$ via \cref{eq:loss:goal,eq:loss:vol,eq:loss:inv,eq:loss:goal_overlap} \label{alg:loss}
    \STATE $\theta_{i+1} \leftarrow \theta_{i} - \text{optimizer.step}(\nabla_\theta \mathcal{L}(\theta_i))$ \label{alg:update}
    \STATE Check if verification criteria are satisfied
\ENDWHILE
\RETURN $\theta$
\end{algorithmic}
\end{algorithm}


\begin{figure*}
     \centering
     \begin{subfigure}[t]{\textwidth}
         \centering
         \includegraphics[clip,trim=250 250 250 0,width=\textwidth]{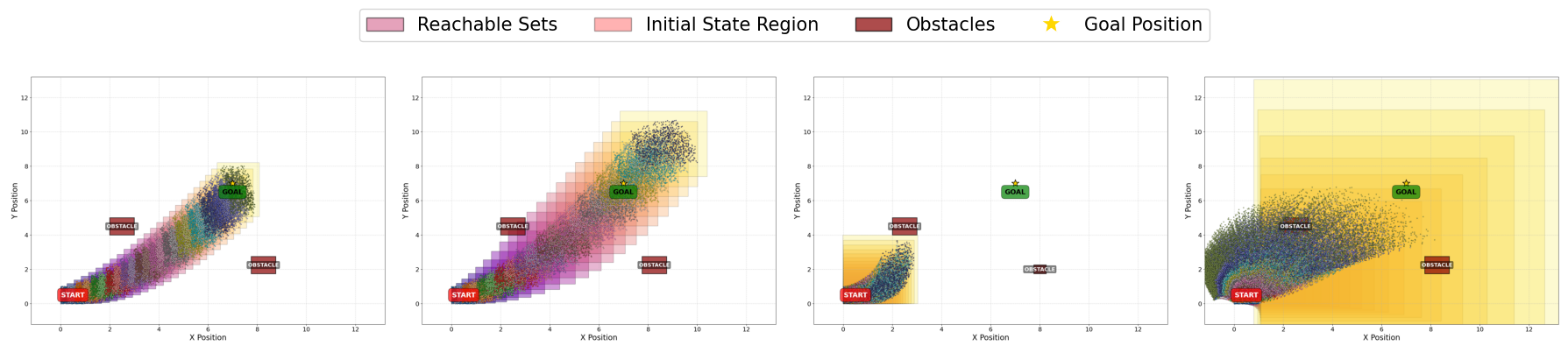}
     \end{subfigure}
     \begin{subfigure}[t]{0.33\textwidth}
         \centering
         \includegraphics[clip,trim=0 0 0 0,width=\textwidth]{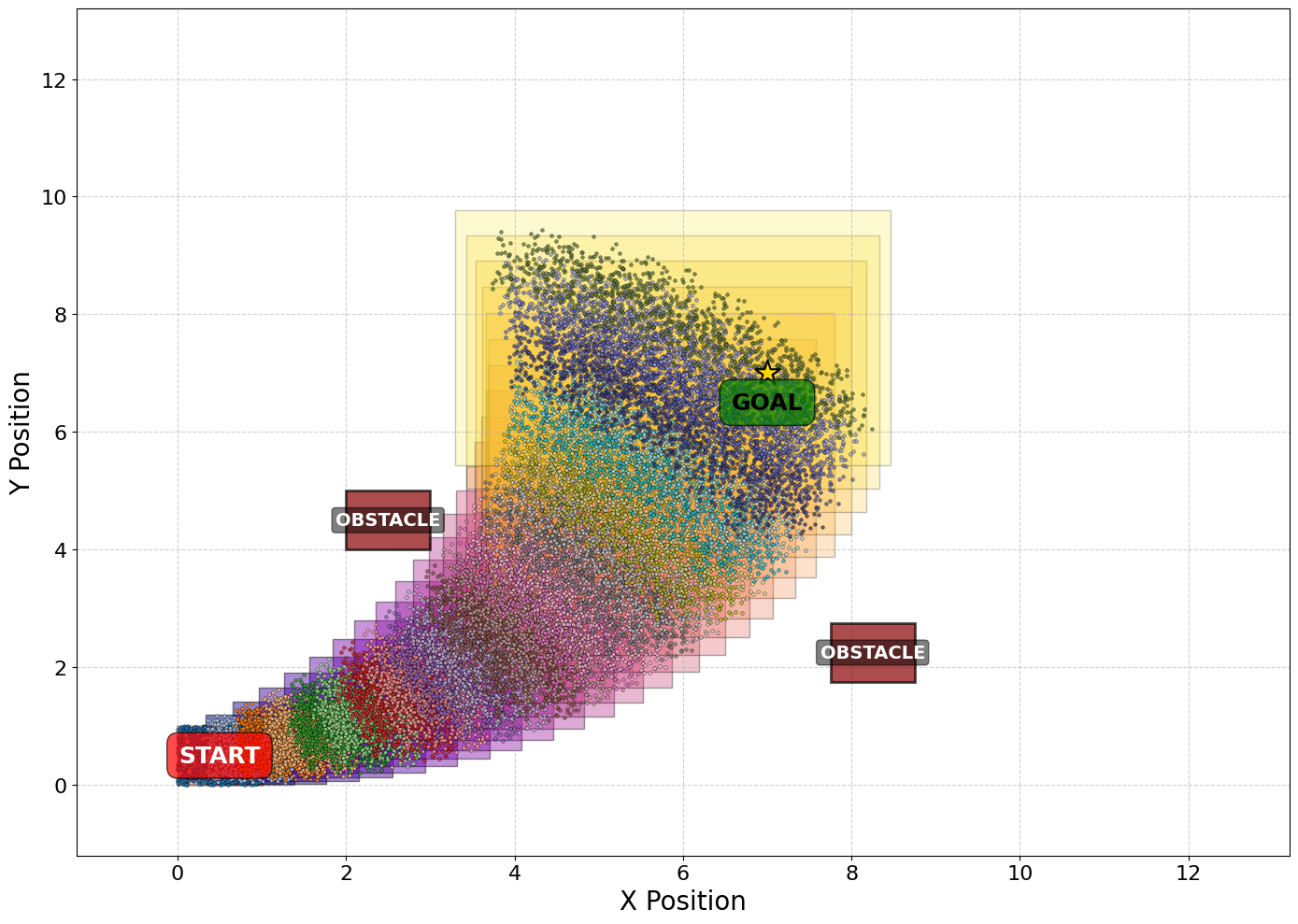}
         \caption{NN Policy, Loss with Numerical Gradients}
         \label{fig:gradient_linear}
     \end{subfigure}%
     \begin{subfigure}[t]{0.33\textwidth}
         \centering
         \includegraphics[clip,trim=0 0 0 0,width=\textwidth]{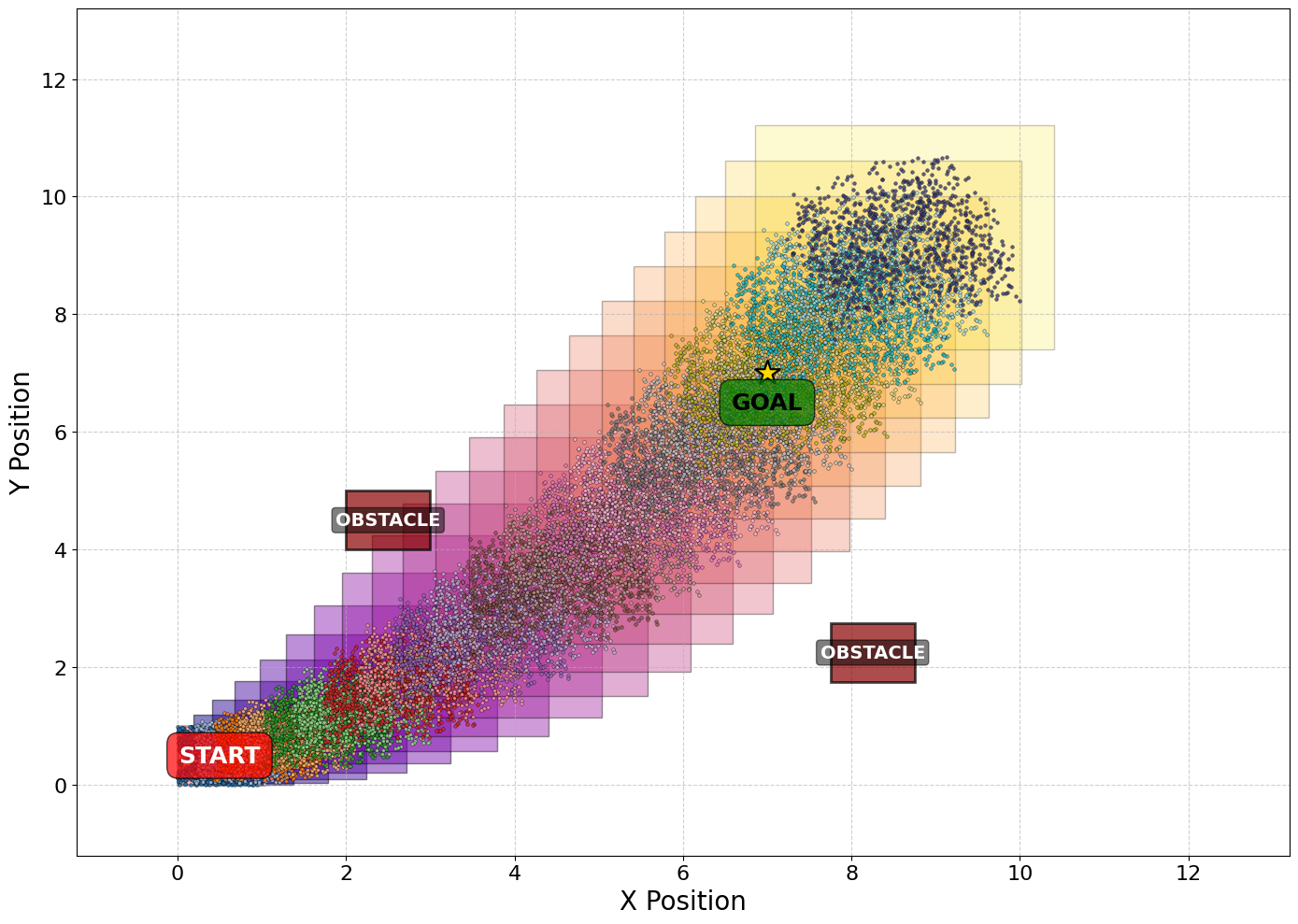}
         \caption{Affine Policy, Loss with CROWN}
         \label{fig:crown_linear}
     \end{subfigure}%
     \begin{subfigure}[t]{0.33\textwidth}
         \centering
         \includegraphics[clip,trim=0 0 0 0,width=\textwidth]{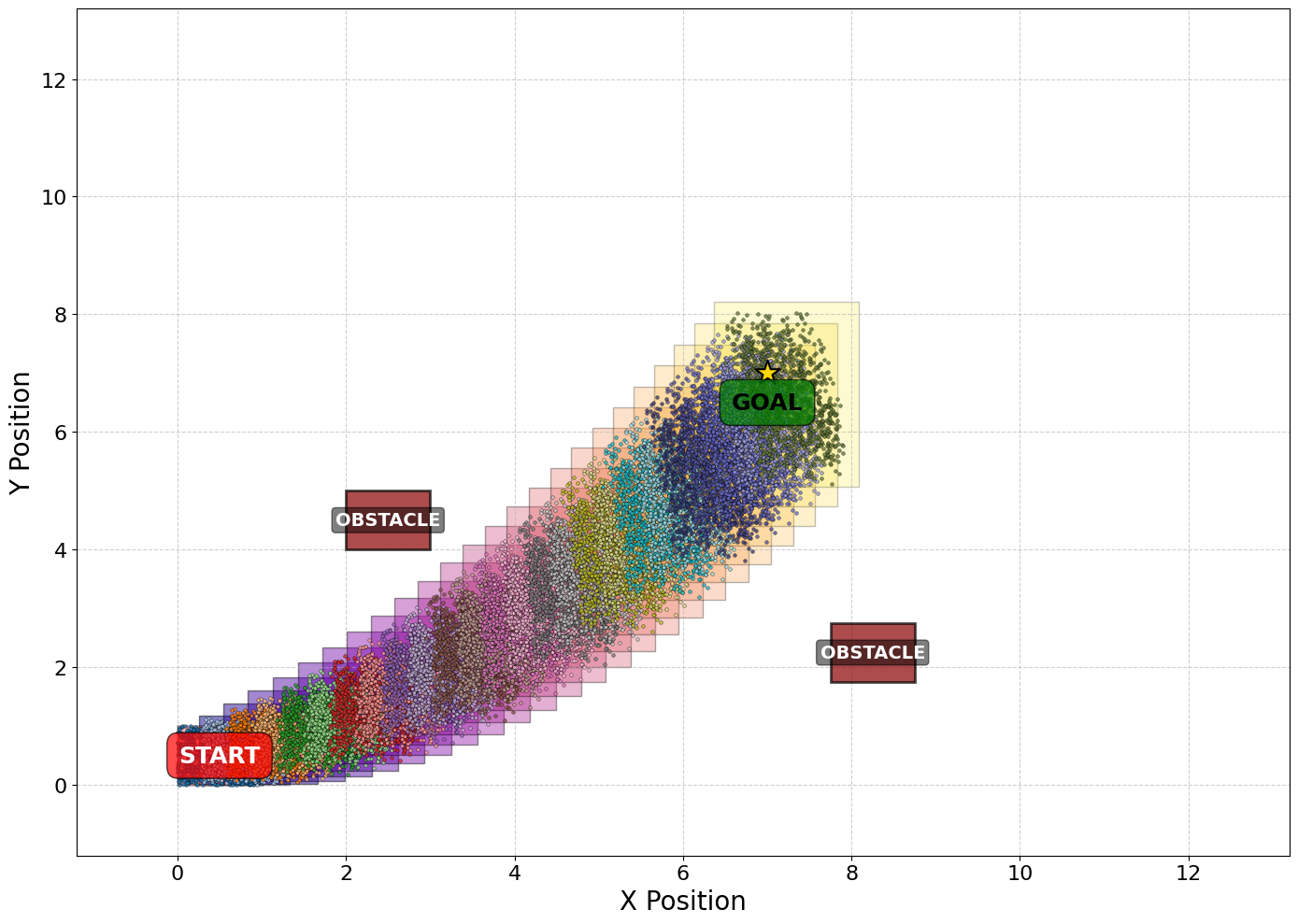}
         \caption{ NN Policy, Loss with CROWN}
         \label{fig:crown_nn}
     \end{subfigure}%
    \caption{Comparison of reachable sets computed by each method for a unicycle model and reach-avoid specification. The policy trained with numerical gradients (left) is sensitive to the initial state. The affine policy trained with CROWN loss (middle) reaches the goal and its sampled trajectories avoid the obstacles, but the reachable set bounds collide with an obstacle. The NN policy (right) trained with CROWN loss reaches the goal, and the reachable set bounds are sufficiently tight that they also avoid the obstacles.}
    \label{fig:all_methods_comparison}
\end{figure*}

The verification-guided training procedure is summarized in Algorithm 1.
After randomly initializing the policy, each training iteration involves computing the reachable set bounds $\bar{\mathcal{R}}_1, \ldots, \bar{\mathcal{R}}_T$ for some time horizon using CROWN~\cite{zhang2018efficient} (\cref{alg:crown}). Then, after calculating the terms of the loss function (\cref{alg:loss}), the optimizer takes a step in the controller parameter space (\cref{alg:update}).
This loop continues until a termination condition is reached (e.g., max number of iterations, convergence).

We note that the use of soft penalties as opposed to hard constraints during training could lead to systems that do not necessarily meet the safety specs.
In those cases, ideas proposed in \cite{wu2024verified}, such as using $n$-step recursive calculations of the forward reachable sets, or training different policies for different subsets of $\mathcal{X}_0$, could be beneficial.
Nonetheless, the experiments in the next section demonstrate various cases where the modified loss does indeed lead to specification satisfaction.







\section{Results}

This section demonstrates the proposed method on several control synthesis tasks with different specifications. First, we show that the differentiable reachability-based approach enables learning a policy for a unicycle to avoid obstacles and reach a goal region, with faster learning and better performance than a prior approach based on numerical gradient estimates~\cite{wang2021verification}. Next, we show that the bound tightness term in the loss function leads to a system whose reachable sets are much tighter, with similar performance on the task, which would enable the use of fast verification methods during runtime. Then, we demonstrate that the invariance term in the loss can lead to an invariant set around the goal region. Finally, we highlight the scalability of the method on a 6D quadrotor model for obstacle avoidance while flying toward a goal region.


\subsection{Obstacle Avoidance: Unicycle}

The discrete-time unicycle system evolves according to:
\begin{equation}
\begin{aligned}
x_{t+1} &= x_t + v_t\cos(\theta_t) \\
y_{t+1} &= y_t + v_t\sin(\theta_t) \\
\theta_{t+1} &= \theta_t + \omega_t,
\end{aligned}
\end{equation}
where state $\mathbf{x}=[x, y, \theta]$ is composed of position and heading angle, and control inputs $\mathbf{u}=[v, \omega]$ are linear and angular velocity, respectively.

\begin{figure*}[t]
    \centering
    \begin{subfigure}[t]{\textwidth}
         \centering
         \includegraphics[clip,trim=300 930 300 0,width=0.8\textwidth]{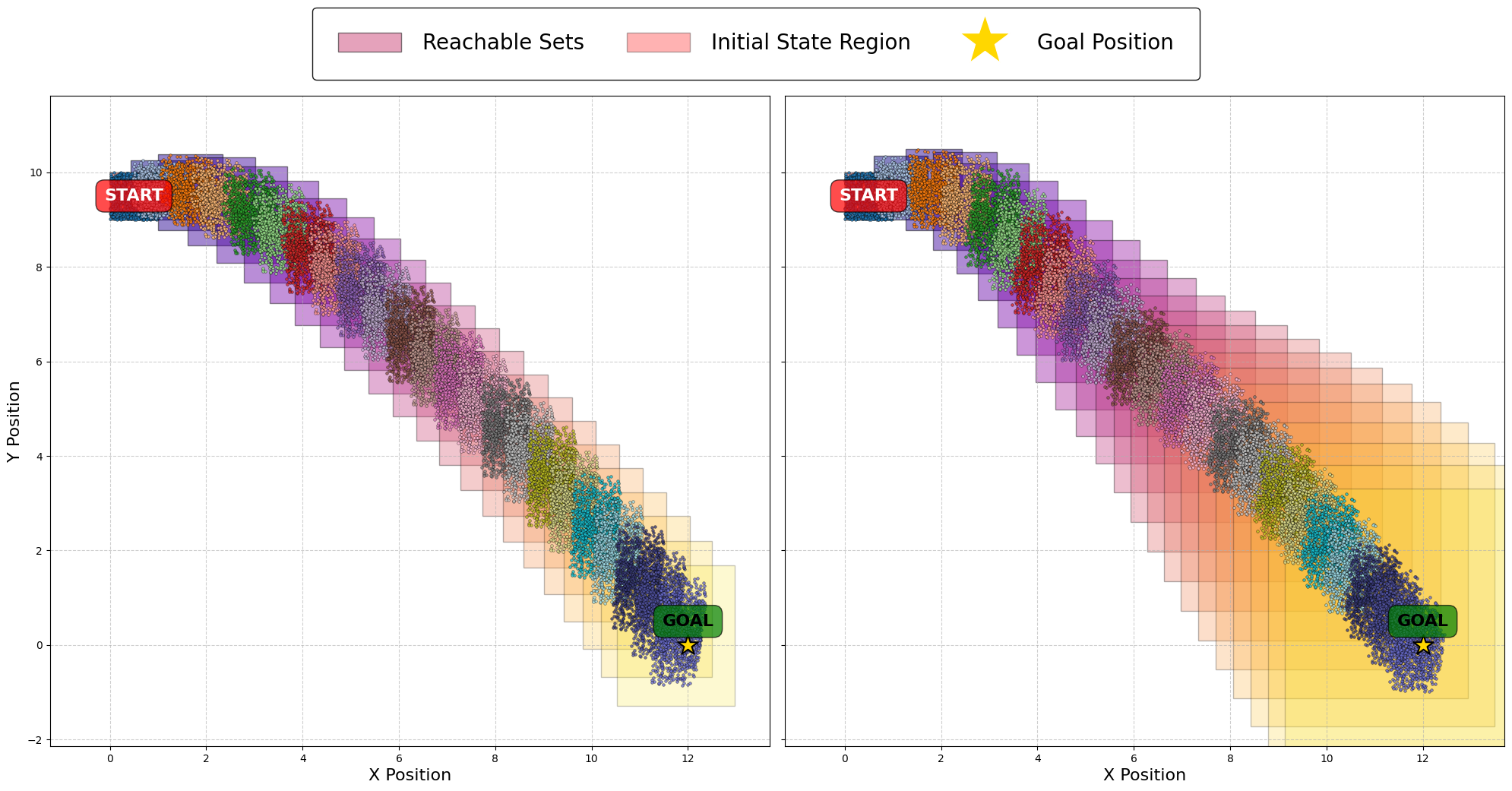}
     \end{subfigure}
     \begin{subfigure}[t]{0.5\textwidth}
         \centering
         \includegraphics[clip,trim=0 0 970 120,width=\textwidth]{figures/unicycle.png}
         \caption{With $w_\text{vol}=4$}
         \label{fig:tightness:tight}
     \end{subfigure}%
     \begin{subfigure}[t]{0.5\textwidth}
         \centering
         \includegraphics[clip,trim=1030 0 0 120,width=0.94\textwidth]{figures/unicycle.png}
         \caption{With $w_\text{vol}=0.05$}
         \label{fig:tightness:loose}
     \end{subfigure}%
     \caption{Effect of Bound Volume Loss, $\mathcal{L}_\text{vol}$. For the unicycle system with a different start/goal from before, the reachable sets calculated by CROWN after training are much tighter when $w_\text{vol}=4$ (left) than $w_\text{vol}=0.05$ (right), even though the system's performance and true reachable sets are nearly identical. This highlights the benefit of including $\mathcal{L}_\text{vol}$ in the training process: it enables obtaining reasonably tight bounds even with a relaxed verification algorithm.}
    \label{fig:tightness}
\end{figure*}

The system must navigate from initial state $\mathbf{x}_0 = (0,1)$ to a goal region $\mathcal{G}$ centered at $(7,7)$, while avoiding two obstacles, $\mathcal{A}_0$ centered at $(2,4)$ and $\mathcal{A}_1$ at $(8,2)$. Initial state uncertainty is bounded by $\pm 0.1$ in each dimension.


\cref{fig:all_methods_comparison} shows the reachable sets after training with three different methods. 
On the left (\cref{fig:gradient_linear}), a NN control policy was trained using the numerical gradient approximation technique from~\cite{wang2021verification} to update the control parameters during training.
While these numerical gradient approaches worked in our experiments replicating the Van der pol oscillator\footnote{Furthermore, for the Van der pol oscillator system (from~\cite{wang2021verification}, not shown here), the training time was 187.60 seconds using the numerical gradient approach~\cite{wang2021verification}, compared to 44.52 seconds with our proposed method.} and cruise control examples from \cite{wang2021verification}, that method did not perform well on this unicycle system for either an affine policy or NN policy (not shown).
Conversely, the proposed method begins to achieve the specification given only an affine policy and performs well with a NN policy.

In the middle (\cref{fig:crown_linear}) is an implementation of the proposed \cref{alg:verification_in_loop} with an affine policy, where $\mathbf{k}\in\mathbb{R}^{3 \times 2}$ and $\mathbf{b}\in\mathbb{R}^2$. The vehicle reaches the goal (and goes past it) while avoiding the obstacles, but the reachable set bounds calculated via CROWN loose enough to intersect the obstacles. This would prevent an obstacle avoidance proof without a more expensive verifier.

On the right, (\cref{fig:crown_nn}) uses a NN control policy. With this policy, the system successfully reaches the goal region in 24 steps, and neither the ``true'' reachable set samples and the over-approximations, $\bar{\mathcal{R}}_1, \ldots, \bar{\mathcal{R}}_{24}$ intersect with the avoid sets. The NN control policy is composed of 3 hidden layers, with [16, 32, 16] neurons, ReLU activations, and parameters initialized using scaled uniform distribution, between $\pm 0.1\sqrt{\frac{6}{n_{\text{in}} + n_{\text{out}}}}$. Other hyperparameters include $T=24$, Adam optimizer with learning rate 1e-4, and 20,000 epochs.

For the CROWN methods, the loss function was
\begin{align}
\mathcal{L}(\theta) = w_\text{goal} \mathcal{L}_\text{goal} + w_\text{overlap\_obs} \mathcal{L}_\text{overlap\_obs} + \nonumber\\ w_\text{overlap\_goal} \mathcal{L}_\text{overlap\_goal} + w_\text{vol} \mathcal{L}_\text{vol},
\end{align}
with $w_\text{goal}=8$, $w_\text{overlap\_danger}=-15$, $w_\text{overlap\_goal}=20$, $w_\text{vol}=0.5$.
$\mathcal{L}_\text{overlap\_goal}$ uses \cref{eq:loss:goal_overlap} as written, and $\mathcal{L}_\text{overlap\_obs}$ replaces the goal center with each obstacle center.
We note that the loss function for the numerical gradient approach was slightly different and will refer readers to the code for exact implementation details.

\subsection{Bound Tightness}

To illustrate the effect of the bound tightness loss term, $\mathcal{L}_\text{vol}$ from \cref{eq:loss:vol}, \cref{fig:tightness} shows a unicycle system controlled by a policy trained with $w_\text{vol}=4$ (left), and $w_\text{vol}=0.05$ (right).
Both trained policies lead to almost identical system behavior and \textit{true} reachable sets according to the sampled trajectories. However, the bounds shown using CROWN, a relaxed/fast verifier, are much tighter after considering the relaxations throughout training. In other words, it would take much more computational effort (e.g., using branch-and-bound~\cite{everett2021reachability}, one-shot analysis~\cite{chen2023one}) to establish bounds of similar tightness for the verification-unaware system.
This result is significant because enabling the use of a lightweight verifier to compute reachable set bounds of reasonable tightness at run-time is important for safety-critical control systems.

The total loss function in this example is:
\begin{align}
\mathcal{L}(\theta) = w_\text{goal} \mathcal{L}_\text{goal} + w_\text{overlap\_goal} \mathcal{L}_\text{overlap\_goal} + w_\text{vol} \mathcal{L}_\text{vol},
\end{align}
with $w_\text{overlap\_goal}=-15$ and $w_\text{goal}=8$.

\begin{figure}[t]
    \centering
    \includegraphics[width=1\linewidth]{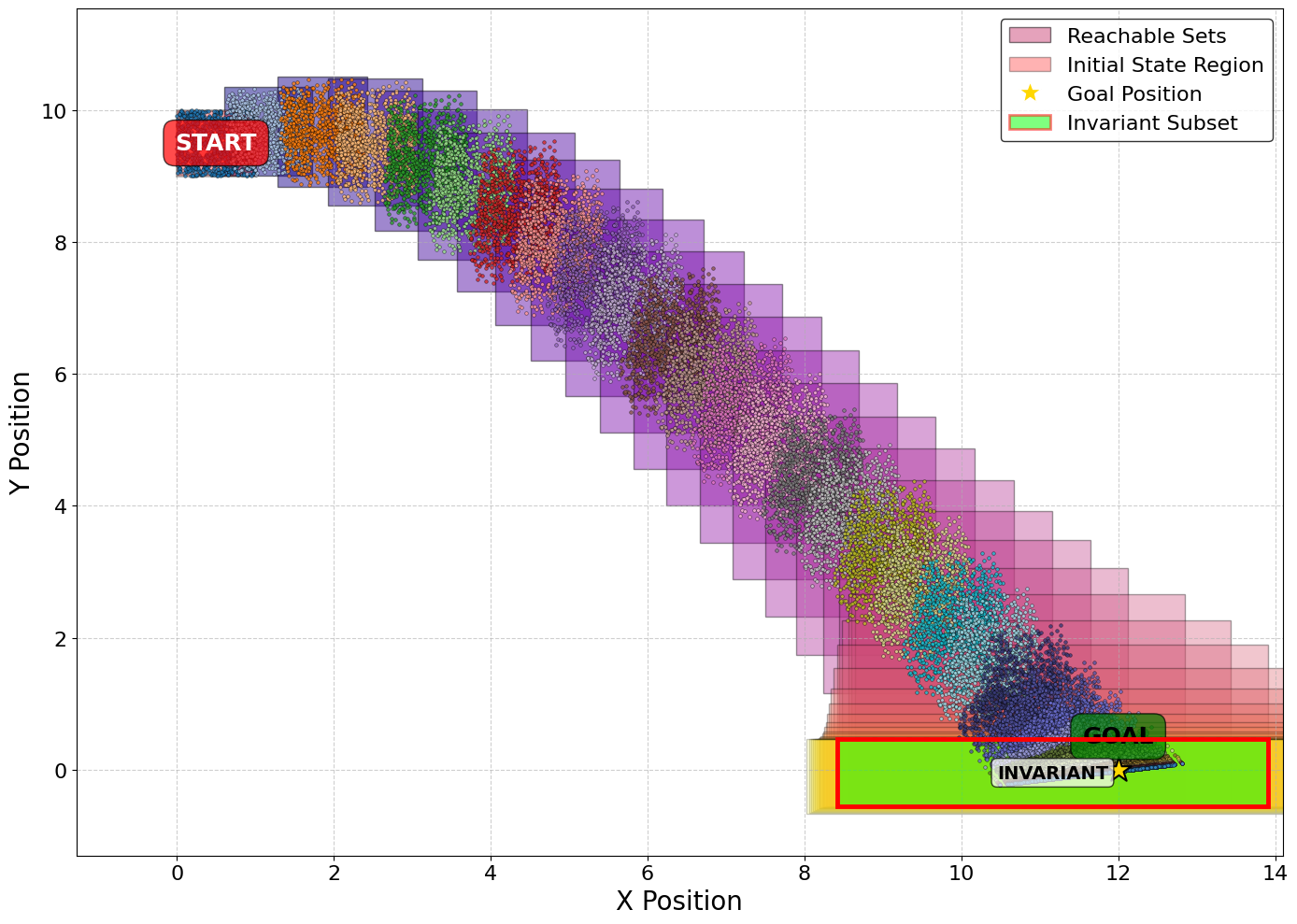}
     \caption{Effect of Invariance Loss, $\mathcal{L}_\text{inv}$. Without considering this loss, the reachable sets often continue to grow or go past the goal (e.g., see \cref{fig:tightness}). Here, with $w_\text{inv}=100$, the reachable set bounds at the 22nd timestep (bright green set) is a forward invariant set, i.e., the system will never leave this set once it enters. This was confirmed by checking that this set's next reachable set (using CROWN) is a subset.}
    \label{fig:unicycle:invariance}
\end{figure}

\subsection{Invariance}

\cref{fig:unicycle:invariance} shows that including $\mathcal{L}_\text{inv}$ from \cref{eq:loss:inv} in the loss function can enable finding an invariant set around the goal. In particular, the algorithm noted that $\bar{\mathcal{R}}_{23} \subseteq \bar{\mathcal{R}}_{22}$. This means that there is no state in $\bar{\mathcal{R}}_{22}$ that could lead to a state \textit{not} in $\bar{\mathcal{R}}_{22}$ in one timestep (and thus forever), which makes $\bar{\mathcal{R}}_{22}$ a forward invariant set.

This experiment used $\mathcal{X}_0=[[0,1],[9,10],[0,\pi/6]]$, obstacles at $(8,8)\pm0.5$ and $(4,4)\pm0.5$, goal at $(12,0)\pm0.5$, $w_\text{goal}=8$, $w_\text{obstacle\_overlap}=20$, $w_\text{goal\_overlap}=15$, $w_\text{inv}=100$, $t_\text{inv}=22$, $T=40$, and trained with learning rate of 1e-4. The loss plateaued at 1,293 after 16,400 epochs, and the invariant set was $x\in[8.417, 13.911]$ and $y\in[-0.553, 0.458]$.

Recent work also investigated finding invariant sets for systems with learned control policies \cite{fazlyab2020safety}. While \cite{fazlyab2020safety} focused on finding the largest possible invariant set for an already trained policy using semidefinite programming (SDP), our approach focuses on training the policy to encourage the existence of an invariant set. Since CROWN is generally faster and scales to larger NNs compared to SDP-based verifiers~\cite{everett2021reachability}, the proposed approach has a key advantage of being able to bias the training process such that only a linear relaxation-based method was needed to find an invariant set.

\subsection{Scalability: 6D Quadrotor}

\begin{figure}[t]
    \centering
    \includegraphics[trim=0 20 0 30,clip,width=1\linewidth]{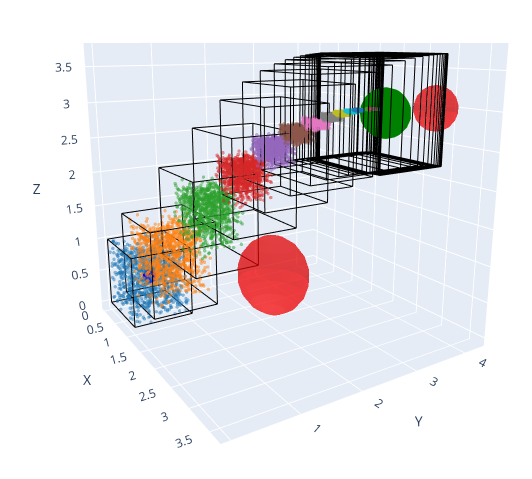}
     \caption{Obstacle avoidance with 6D quadrotor model ($xyz$ positions shown). The trained policy starts in the set with blue samples and reaches the goal (green) in 20 timesteps. Along the way, both the true system trajectories (samples) and reachable set bounds (calculated with CROWN) avoid the obstacles.}
    \label{fig:quadrotor_obstacles}
\end{figure}

To demonstrate the scalability of the method, \cref{fig:quadrotor_obstacles} shows the reachable sets in $(x,y,z)$ position for a 6D quadrotor model. More specifically, this is a model with 3 double integrators that includes drag and coupling between directions,
\begin{align}
    x_{t+1} &= x_t + v_{x,t}\Delta t + (a_{x,t} + c_c v_{y,t}v_{z,t}-c_d v_{x,t} \lvert v_{x,t} \rvert)\frac{\Delta t^2}{2} \nonumber\\
    y_{t+1} &= y_t + v_{y,t}\Delta t + (a_{y,t} + c_c v_{z,t}v_{x,t}-c_d v_{y,t} \lvert v_{y,t} \rvert)\frac{\Delta t^2}{2} \nonumber\\
    z_{t+1} &= z_t + v_{z,t}\Delta t + (a_{z,t} + c_c v_{x,t}v_{y,t}-c_d v_{z,t} \lvert v_{z,t} \rvert)\frac{\Delta t^2}{2} \nonumber\\
    v_{x,t+1} &= v_{x,t}+(a_{x,t} + c_c v_{y,t}v_{z,t}-c_d v_{x,t} \lvert v_{x,t} \rvert)\Delta t \nonumber\\
    v_{y,t+1} &= v_{y,t}+(a_{y,t} + c_c v_{x,t}v_{z,t}-c_d v_{y,t} \lvert v_{y,t} \rvert)\Delta t \\
    v_{z,t+1} &= v_{z,t}+(a_{z,t} + c_c v_{x,t}v_{y,t}-c_d v_{z,t} \lvert v_{z,t} \rvert)\Delta t, \nonumber
\end{align}
with $\Delta t = 0.4$, $c_d=0.01$, $c_c=0.005$, and let $\mathbf{p}=[x, y, z], \mathbf{v}=[v_x, v_y, v_z], \mathbf{x}=[\mathbf{p}, \mathbf{v}], \mathbf{u}=[a_x, a_y, a_z]$.

Other parameters include initial state set $\mathcal{X}_0=[[0,1],[0,1],[0,1],[-0.5,0.5],[-0.5,0.5],[-0.5,0.5]]$, goal point $(3,3,3)$, one obstacle at $[2.5, 1.5, 1.0]$ with radius 0.5, another obstacle at $[3.0, 4.0, 3.0]$ with radius 0.3, and $T=20$. The NN controller with hidden layer sizes $[24, 48, 24]$ was trained for 3,000 epochs to a loss of 2,382 using Adam with learning rate 3e-3.

For this system, the full loss function was,
\begin{align}
    \mathcal{L}_\text{quad}(\theta) = &w_\text{goal} \mathcal{L}_\text{goal} + w_\text{overlap} \mathcal{L}_\text{overlap} + w_\text{vel} \mathcal{L}_\text{vel} + w_\text{vol} \mathcal{L}_\text{vol} + \nonumber\\ &\quad w_\text{obs\_entry} \mathcal{L}_\text{obs\_entry} + w_\text{obs\_prox} \mathcal{L}_\text{obs\_prox},
\end{align}
with weights $w_\text{goal}=50$, $w_\text{overlap}=-50$, $w_\text{vel}=0.05$, $w_\text{vol}=40$, $w_\text{obs\_entry}=500$, and $w_\text{obs\_prox}=100$.
Along with $\mathcal{L}_\text{goal}$ from \cref{eq:loss:goal}, $\mathcal{L}_\text{overlap}$ from \cref{eq:loss:goal_overlap}, and $\mathcal{L}_\text{vol}$ from \cref{eq:loss:vol}, this experiment additionally penalized non-zero velocities:
\begin{align}
    \mathcal{L}_\text{vel}&=\sum_{t=0}^{T} \lvert\bar{\mathbf{v}}_{t}\rvert + \lvert\ubar{\mathbf{v}}_{t}\rvert.
\end{align}
The volume loss also included a constant bias of $-\text{vol}(\mathcal{X}_0)$ at each timestep, and was scaled by $\frac{1}{\text{vol}(\mathcal{X}_0)}$, but these constants should be able to be removed.

For the obstacle loss terms, let $\mathbf{n}_{t,j}=\text{max}(\ubar{x}_t, \text{min}(\bar{x}_t, \mathbf{o}_j))$ be the nearest point on $\bar{\mathcal{R}_t}$ to the center of the $j$-th obstacle, $\mathbf{o}_j$, whose radius is $r_j$. Then,
\begin{align}
    \mathcal{L}_\text{obs\_entry}=\sum_{t=0}^{T} \text{max}(r_j - \|\mathbf{n}_{t,j}-\mathbf{o}_j\|_2, 0)^2
\end{align}
penalizes reachable sets within an obstacle's radius, and 
\begin{align}
\mathcal{L}_\text{obs\_prox}=\sum_{t=0}^T \sum_{j=0}^{A} \text{max}(r_j + m - \|\mathbf{n}_{t,j}-\mathbf{o}_j\|_2, 0)^2
\end{align}
penalizes reachable sets that get closer than a distance $m=1$ to each obstacle.

Across the 20 timesteps in the trajectory, none of the reachable sets intersect with the obstacles and the samples move toward the goal region (but we did not optimize for invariance here).

We hypothesize that the training process used here would make relaxed forward and backward reachability approaches less reliant on partitioning schemes, as in~\cite{rober2023backward}, which can be difficult to scale to higher dimensional systems. However, we leave that analysis for future work.

\section{Conclusion}
This paper proposed a new approach for learning NN control policies based on safety specifications.
In particular, the approach uses CROWN\cite{zhang2018efficient} to calculate reachable sets at each training iteration, then employs loss terms to encourage the system to robustly achieve the safety specs.
Numerical experiments on a quadrotor model and unicycle model highlight the ability of this approach to lead to learned control policies that satisfy desired reach-avoid and invariance specifications.
Future work will investigate further scalability to larger systems, specifications for more challenging objectives, and integration with imitation learning for safely learning from expert demonstrations.


\balance
\bibliographystyle{IEEEtran}
\bibliography{bibliography}

\end{document}